\documentclass[prl,floatfix,amsmath,twocolumn]{revtex4}
\usepackage{graphicx}
\usepackage[colorlinks]{hyperref} 
\usepackage{amssymb}
\usepackage{mathptm,times,amsmath}
\usepackage{wasysym}
\usepackage{psfrag}
\usepackage{bm}
\usepackage{color}

\newcommand{\ket}[1]{|#1\rangle}
\newcommand{\bra}[1]{\langle#1|}

\newcommand{\bl}{\begin{itemize}}
\newcommand{\el}{\end{itemize}}

\newcommand{\M}[1]{\ket{\sf m#1}}

\newcommand{\beq}{\begin{equation}}
\newcommand{\eeq}{\end{equation}}
\newcommand{\beqa}{\begin{eqnarray}}
\newcommand{\eeqa}{\end{eqnarray}}

\usepackage{graphicx}
\usepackage[mathscr]{eucal}
\usepackage{amsmath}
\usepackage{amssymb}
\usepackage{epstopdf}
\usepackage{epsfig}
\begin{document}

\title{Non-locality of non-Abelian anyons}

\author{G.K. Brennen$^{1}$}
\author{S. Iblisdir$^{2}$}
\author{J.K. Pachos$^{3}$}
\author{J.K. Slingerland$^{4,5}$}
\affiliation{$^1$ Centre for Quantum Information Science and Security, Macquarie
  University, 2109, NSW Australia}
\affiliation{$^2$Dept. Estructura i Constituents de la Mat$\grave{e}$ria, Universitat de Barcelona, 08028 Barcelona, Spain}
\affiliation{$^3$School of Physics and Astronomy, University of Leeds, Leeds LS2 9JT, UK}
\affiliation{$^4$Dublin Institute for Advanced Studies, School of Theoretical Physics, 10 Burlington Rd, Dublin, Ireland}
\affiliation{$^5$Department of Mathematical Physics, National University of Ireland, Maynooth, Ireland}


\begin{abstract}

Topological systems, such as fractional quantum Hall liquids, promise to
successfully combat environmental decoherence while performing quantum
computation. These highly correlated systems can support non-Abelian anyonic
quasiparticles that can encode exotic entangled states. To reveal the
non-local character of these encoded states we demonstrate the violation of suitable
Bell inequalities. We provide an explicit recipe for the preparation,
manipulation and measurement of the desired correlations for a large class
of topological models. This proposal gives an operational measure of
non-locality for anyonic states and it opens up the possibility to violate
the Bell inequalities in quantum Hall liquids or spin lattices.

\end{abstract}

\pacs{...}

\maketitle

Quantum mechanics is a non-local theory: it allows for correlations between
distant systems that cannot be explained in terms of a local preparation.
Many believed that non-locality was due to incompleteness of quantum theory.
Einstein, Podolsky and Rosen (EPR) in their seminal work~\cite{EPR}
aimed to resolve this by introducing~\emph{local hidden} variables. Their
values would complement the information supplied by quantum mechanics, thus
restoring locality. Bell inequalities aim at validating or rejecting this
view from experimental data~\cite{Bell}. To date, unlike local hidden
variable (LHV) theories, the predictions of quantum mechanics have been
consistent with all Bell tests. 

Particle exchange gives a striking example of non-locality in quantum mechanics. For bosons and fermions, one can deal with the exchange interactions by imposing non-local constraints on the form of the wave function.  More generally, the wave function can transform in a nontrivial representation of the fundamental group of configuration space when particles are adiabatically exchanged~\cite{Leinaas77}. For planar systems, this group is the braid group and particles transforming in nontrivial braid group representations have been dubbed {\it anyons}~\cite{wilczek}. Anyonic exchange interactions are topological in nature and do not change on variation of the distance between the particles, or the metric of the spacetime manifold. One mechanism for such interactions is the Aharonov Bohm effect~\cite{AB}.

It is natural to ask if the non-local correlations of anyons can in principle be explained using local hidden variables. In this paper we answer this question for a number of anyon models, by constructing a Bell test for anyonic degrees of freedom and showing violation of the associated Bell inequalities.  


Anyons can be split into two main categories, they can be Abelian or non-Abelian.
Given labels $\{a_j\}$ for the different types of anyons we assign fusion
rules that determine the outcome of bringing two anyons together, $a_i\times
a_j=\sum_k N_{a_ia_j}^{a_k}a_k$.  Here $N_{ab}^c\in~\mathbb{N}$ counts the 
number of ways of combining $a$ and $b$ to obtain $c$. Non-Abelian anyons
have $\sum_c N_{ab}^c\geq 2$ for some pair $a,b$, while in the Abelian case, the labels of the fused anyons determine a unique outcome which can be determined in a unique way. 

In a physical system with anyons, the low energy part of  the Hilbert space can be thought of as a tensor product $\mathcal{H}=\mathcal{H}_{\rm local}\otimes\mathcal{H}_{\rm non-local}$, where the first factor describes local degrees of freedom, which we will ignore, and the second describes topological degrees of freedom associated with the anyons. These topological degrees of freedom may arise as a result of nontrivial topology of the space supporting the anyons. For Abelian anyons, this is in fact the only possibility; in the Abelian toric code models \cite{Kitaev} for instance, the non-local degrees of freedom are described by elements of the first homology groups of the surface with finite group coefficients. In principle, one probe non-local correlations in these topological degrees of freedom, but the observables involved would need to be non-local themselves~\cite{foot}. 

For non-Abelian anyons, even on a contractible surface, there are non-local degrees of freedom associated with the different fusion outcomes. A number of proposals have been made on how the associated quantum numbers, or topological charges, might be measured by interferometry, see \cite{DasSarma05}, \cite{TQC} for further references and \cite{Bonderson08} for an overview of the measurement theory. We will not go into the details of interferometric measurements here, but rather just assume that we can do projective measurements onto the various fusion channels.
The non-local Hilbert state space of $n$ anyons $(a_1,a_2,\ldots a_{n})$ with total charge $c$ has ${\rm
dim}(\mathcal{H}_{\rm non-local})=\sum_{b_1,b_2,\ldots,
b_{n-2}}N_{a_1a_2}^{b_1}N_{b_1a_3}^{b_2}N_{b_2a_4}^{b_3}\cdots
N_{b_{n-2}a_n}^{c}$.
This Hilbert space usually does not admit a tensor product structure, e.g.~the dimension could be
prime, and thus does not obviously fit the usual paradigm for tests of non-locality.
Nevertheless, we show that topological interactions can indeed be
used to demonstrate non-locality in the EPR sense.
In order to do this, we consider two
classes of anyonic theories: the $SU(2)_k$ models, including the Fibonacci
model~\cite{TQC}, and a model based on discrete gauge theory~\cite{Bais:92,Propitius98}.   These models are important both for their potential to process quantum information fault tolerantly and for their viability for experimental realization.  For these
cases the fusion spaces are at most one dimensional, i.e.~$N_{ab}^c< 2$ for all $(a,b,c)$.  
Non-commuting measurements project onto different ways of combining
particles $a,b,c$ to yield $d$.  Measurement bases are labelled by the intermediate
products $x$ and $x'$ obtained by fusing $a,b,c$ and are related by the recoupling formula:
$\ket{(ab)c\rightarrow d;x}=\sum_{x'}(F_{abc}^d)_x^{x'}\ket{(a(bc)\rightarrow d;x'}$.

To date, most Bell tests have been performed on entangled light beams, but there is certainly an interest in showing that material media can be used to demonstrate non-locality. Experiments involving a photon and an atom, two atoms, or even kaons have been proposed or even carried out~\cite{monroe}. The schemes we are proposing contribute to this effort of using ever new media for Bell tests. As we shall see, some of them could be implemented in a fractional quantum Hall liquid~\cite{TQC}, while others can be associated with arrays of Josephson junctions~\cite{Doucot} or atoms in optical lattices~\cite{Aguado}.

\begin{figure}
\begin{center}
\includegraphics[width=6cm]{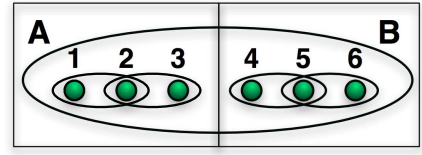}
\caption{\label{fig:1}
A Bell type measurement on six particles. First a joint measurement (large
oval) on all six particles is made and the result kept if the total charge
(or spin) is zero. Alice performs measurements of total charge on pairs
$1,2$ and $2,3$ and Bob performs measurements on pairs $4,5$ and $5,6$. For
some quantum states the correlator $\langle W\rangle$ exceeds the bound set
by local hidden variable theories.}
\end{center}
\end{figure}

 In order to build intuition for the anyonic case
we describe the general framework by employing distinguishable spin-$1/2$ particles 
with non-trivial fusion properties. The fusion
rules are interpreted as the angular momentum decomposition of tensor
products of vector spaces.
Consider a system divided into two spatially non-overlapping subsystems $A$
and $B$, conveniently labeled as Alice and Bob, each one possessing three
spin-$1/2$ particles, as seen in Fig.~\ref{fig:1}. First, we
perform a joint measurement on the total spin $\vec{S}_{\rm tot}=\sum_j
\vec{s}_j$ and post-select the $S_{\rm tot}=0$ outcome that has state space
dimension five. Second, we define a set of measurement operators
$\{\Upsilon^A_{1,2},\Upsilon^A_{2,3},\Upsilon^B_{4,5},\Upsilon^B_{5,6}\}$,
where $\Upsilon_{i,j}=(\vec{s}_i+\vec{s}_j)^2-{\bf 1}$. The eigenvalues of
$\Upsilon_{i,j}$ are $+1$ in the triplet space and $-1$ for the singlet and
the operators $\Upsilon^{A(B)}$ act on the subsystems $A(B)$.  The operator
pair within $A$ or $B$ is non-commuting but
$[\Upsilon_{i,j}^A,\Upsilon_{j,k}^B]=0$.  Consider the expectation value of
the operator
\begin{equation}
W=\Upsilon^A_{1,2}\Upsilon^B_{4,5} + \Upsilon^A_{1,2}
\Upsilon^B_{5,6}-\Upsilon^A_{2,3}\Upsilon^B_{5,6} +
\Upsilon^A_{2,3}\Upsilon^B_{4,5}.
\label{W}
\end{equation}
For a classical theory, even in the presence of local hidden variables
(LHV)~\cite{localhidden}, the Bell inequality for W is $|\langle
W\rangle_{\rm LHV}|\leq2$. This can be derived straightforwardly as follows~\cite{CHSH}. 
Assume independence of the two subsystems (locality) so that the joint probabilities for pairs of outcomes is just the product of the individual probabilities which could depend on a hidden variable $\lambda$, drawn from a fixed distribution $p(\lambda)$. For the above quorum of observables with outcomes $\{m^A_{1,2},m^A_{2,3},m^B_{4,5},m^B_{5,6}\}\in \pm 1$ we have
\[
(m^A_{2,3}+m^A_{1,2})m^B_{4,5}-(m^A_{2,3}-m^A_{1,2})m^B_{5,6}=\pm 2.
\]
Hence, in the LHV model, the outcomes must satisfy
\[
\begin{array}{lll}
|W_{\rm LHV}|&=&|\int d\lambda p(\lambda)\langle (\Upsilon^A_{1,2}(\lambda)\Upsilon^B_{4,5}(\lambda) \\
&&+\Upsilon^A_{1,2}(\lambda)\Upsilon^B_{5,6}(\lambda)-\Upsilon^A_{2,3}(\lambda)\Upsilon^B_{5,6}(\lambda) \\
&&+\Upsilon^A_{2,3}(\lambda)\Upsilon^B_{4,5}(\lambda))\rangle|\\
&=&|\int d\lambda p(\lambda) (m^A_{2,3}(\lambda)+m^A_{1,2}(\lambda))m^B_{4,5}(\lambda)\\
&&-(m^A_{2,3}(\lambda)-m^A_{1,2}(\lambda))m^B_{5,6}(\lambda)|\\
&\leq&2
\end{array}
\]
 Quantum mechanically, the maximum value of $|\langle W\rangle|$ is obtained for eigenstates of $W$ with maximum eigenvalue, i.e. $|\langle W\rangle |\leq \sqrt{7}$. For arbitrary operators in Eq.~(\ref{W}) that have the same commutation structure and square to ${\bf 1}$, quantum mechanics satisfies Tsirelson's inequality~\cite{Cirelson}, $|\langle W\rangle|\leq \sqrt{8}$. Our aim
is to find a violation of the classical upper bound in the subspace of
states with $S_{\rm tot}=0$.  Note that our protocol allows for measuring correlations without the need of a shared reference frame between Alice and Bob~\cite{BRS} thus giving a simple and unambiguous test of Bell inequalities. In the anyonic case treated below the operators $\Upsilon_{i,j}$ also have eigenvalue $-1$ when the fusion outcome is the vacuum and $+1$ otherwise.  

\begin{figure}
\begin{center}
\includegraphics[scale=0.5]{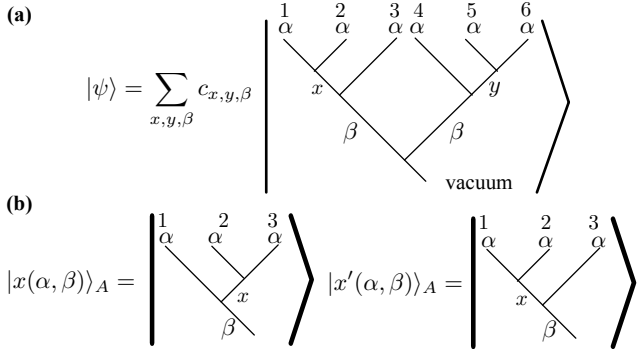}
\caption{\label{fig:3} The state space of anyons in our protocol represented as fusion trees.
(a) An arbitrary state of six $\alpha$ type anyons with trivial total charge
expanded in terms of fusion outcomes local to $A$ and $B$.  In the models considered here, $\beta$ is its own antiparticle, but it is straightforward to generalize. (b) A ``local"
fusion basis satisfying $\Upsilon^A_{2,3}\ket{x(\alpha,\beta)}_A=\pm
\ket{x(\alpha,\beta)}_A$ and $\Upsilon^A_{1,2}\ket{x'(\alpha,\beta)}_A=\pm
\ket{x'(\alpha,\beta)}_A$ for $x$ a vacuum state or a particle  and
similarly for $B$.  For each subsystem the bases are related by an $F$ move:
$\ket{x'(\alpha,\beta)}=\sum_x (F_{\alpha\alpha\alpha}^{\beta})^x_{x'}
\ket{x(\alpha,\beta)}$.}
\end{center}
\end{figure}

There are two natural orthonormal bases for a three particle system based on the two different orders of fusing the three particles. These are graphically represented by fusion trees in Fig.~\ref{fig:3}b. 
The unitary transformation that describes the change from one of these bases to the other is given by the so called $F$ matrices. The basis change for three particles with charges $(a,b,c)$ fusing to $d$ is given by a matrix denoted $F^{d}_{a,b,c}$. For the case of $SU(2)$ these matrices just describe angular momentum recoupling and their matrix elements are the Wigner $6$-$j$ symbols.
For six particles with total spin $0$, we get four natural product bases from the two pairs of bases for each triple. The contributing particle labels are
the spin values $\{0,\frac{1}{2},1,\frac{3}{2}\}$ and the only relevant $F$-matrix for changing between bases is $F_{\frac{1}{2}\frac{1}{2}\frac{1}{2}}^{\frac{1}{2}}$, which, for the $SU(2)$ case is given by
\[
F\equiv F_{\frac{1}{2}\frac{1}{2}\frac{1}{2}}^{\frac{1}{2}}=
\frac{1}{2}\left(
\begin{array}{cc}
1 & \sqrt{3}\\
\sqrt{3} & -1
\end{array}
\right)
\]
in the basis given by fusion trees with intermediate spins $0$ and $1$. We find the following set of orthonormal states for the six particle system, in the local fusion basis as defined in Fig.~\ref{fig:3},
\begin{equation}
\begin{array}{lll}
\ket{\phi_0}&=&\ket{0'}_A\ket{0}_B,\ \
\ket{\phi_1}=\ket{1'}_A\ket{0}_B,\ \
\ket{\phi_2}=\ket{0'}_A\ket{1}_B,\\
\ket{\phi_3}&=&\ket{1'}_A\ket{1}_B,\ \
\ket{\phi_4}=\ket{1(\frac{1}{2},\frac{3}{2})}_A
\ket{1(\frac{1}{2},\frac{3}{2})}_B,
\label{vects}
\end{array}
\end{equation}
where $\ket{x}\equiv\ket{x(\frac{1}{2},\frac{1}{2})}$ and $\ket{x'}=\sum_{x} F_{x'}^x\ket{x}$.
In spin components we have $\ket{\phi_0}=\ket{\Psi^-}_{1,2}\otimes
\ket{\Psi^-}_{3,4}\otimes\ket{\Psi^-}_{5,6}$, where
$\ket{\Psi^-}=(\ket{\uparrow\downarrow}-\ket{\downarrow\uparrow})/\sqrt{2}$, so the state $\ket{\phi_0}$ has three adjacent singlet pairs.   
Notice that in order to have trivial total charge the local bases occur in
pairs that share the same label $\beta$ as defined in Fig. ~\ref{fig:3}.

\begin{figure}[h]
\begin{center}
\includegraphics[scale=0.4]{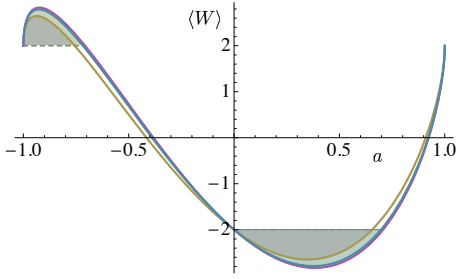}
\caption{\label{fig:4} The expectation value of the Bell witness $W$
as a function of the amplitude of mixing for the  total charge zero states
$\ket{r(a)}$ in Eqs.~(\ref{Eq:Wviolate}).  The yellow, blue, and red lines
correspond to an $SU(2)$, $SO(3)_3$, $SU(2)_2$  theory with six spin-$1/2$,
$\sigma$, $\tau$ particles, respectively. The shaded region corresponds to
states which violate the inequality derived for local hidden variable
models.}
\end{center}
\end{figure}

In the basis
$\{\ket{0}\ket{0},\ket{0}\ket{1},\ket{1}\ket{0},\ket{1}\ket{1}\}\sqcup
\ket{\phi_4}$, we have
\begin{eqnarray}
W=&&\!\!\!\!\!\!\!\!(F^{\dagger}\sigma^zF\otimes
F^{\dagger}\sigma^zF+F^{\dagger}\sigma^zF\otimes \sigma^z+\sigma^z\otimes
F^{\dagger}\sigma^zF-\sigma^z\otimes \sigma^z)
\nonumber \\
&&\oplus 2\ket{\phi_4}\bra{\phi_4}
\label{Wexpand}
\end{eqnarray}
The Hilbert space splits into different sectors labeled by $\beta$, which are conserved by the action of $W$. There is a four dimensional sector with $\beta=1/2$ and a one dimensional sector with $\beta=3/2$, containing $\ket{\phi_4}$. No Bell violation can occur in the $\beta=3/2$ sector, since the measurement operators commute in that sector. Maximally Bell violating states are thus orthogonal to $\ket{\phi_4}$. For the one parameter family of states
\begin{equation}
\ket{r(a)}=\frac{a}{\sqrt{2}}(\ket{\phi_0}+\ket{\phi_3})+
\frac{\sqrt{1-a^2}}{\sqrt{2}}(\ket{\phi_1}-\ket{\phi_2}),
\label{Eq:Wviolate}
\end{equation}
with $(-1\leq a\leq 1)$ we plot the expectation value $\langle W\rangle$
seen in Fig.~\ref{fig:4}. The maximal violation ($\langle W\rangle_{\rm
max,min} =\pm \sqrt{7}\approx \pm 2.6458)$ is obtained for $a_{\rm \pm}=\mp
\sqrt{(7\pm 2\sqrt{7})/14}$.


Consider now a two dimensional system with
quasiparticle excitations described by $SU(2)_k$ Chern-Simons-Witten
theories. The corresponding fusion rules satisfy the addition of angular
momentum with the constraints $j_1\times j_2 \rightarrow j$ only if
$j_1,j_2,j\leq k/2$ and $j_1+j_2+j\leq k$. It is quickly verified that for
$k\geq 3$, the total charge $zero$ sector of six particles labeled by
spin $1/2$ again has five states, labeled by the same fusion trees as in the $SU(2)$ case. 
The $F$ matrices will differ, but for our purposes, the only relevant recoupling is still
$\tilde F_{\frac{1}{2}\frac{1}{2}\frac{1}{2}}^{\frac{1}{2}}$. Computing the
quantum $6$-$j$ symbols, we find (see for instance~\cite{Slingerland:02})
\[
F\equiv
\tilde F_{\frac{1}{2}\frac{1}{2}\frac{1}{2}}^{\frac{1}{2}}=\frac{1}{[2]_q}
\left(
\begin{array}{cc}
1& \sqrt{[3]_q} \\
\sqrt{[3]_q} &-1
\end{array}
\right)
\]
where the quantum integers are defined as
$[m]_q=\frac{q^{m/2}-q^{-m/2}}{q^{1/2}-q^{-1/2}}$ for $m$ integer.  For the
$SU(2)_k$ theories, $q=e^{\frac{2\pi i}{k+2}}$. In the limit $k\rightarrow
\infty$, then $[m]_q\rightarrow m$. As before, we can label the states in the
local fusion basis, as in Eq. \ref{vects}, but with the appropriate $F$
matrix.
The state $\ket{\phi_0}$ is obtained by creating spin-$1/2$ particle
anti-particle pairs at positions $(1,2),(3,4),(5,6)$ out of the vacuum.  For the one
parameter family of states $\ket{r(a)}$ we find maximal violation at
\begin{equation}
\begin{array}{lll}
a_{\rm +}&=&-\frac{1}{\sqrt{8 \cos(\frac{2\pi}{k+2})+
\cos(\frac{4\pi}{k+2})+5}}\Bigg[\cos^2(\frac{2\pi}{k+2})+4\cos(\frac{2\pi}{k+2})\\
&+&\sqrt{2\cos^4(\frac{\pi}{k+2})(8\cos(\frac{2\pi}{k+2})+
\cos(\frac{4\pi}{k+2})+5)}+2\Bigg]^{\frac{1}{2}}\\
\label{as}
\end{array}
\end{equation}
and at $a_-=\sqrt{1-a_+^2}$, where
\[
\langle W\rangle =\pm \sec^2\Big(\frac{\pi}{k+2}\Big)
\sqrt{4\cos\Big(\frac{2\pi}{k+2}\Big)+\frac{1}{2}\cos
\Big(\frac{4\pi}{k+2}\Big)+\frac{5}{2}}.
\]
It is possible to verify that $k\rightarrow\infty$ corresponds to $SU(2)$. A qualitative difference between anyonic systems and the spin systems discussed before is that, while the $z$-components of the spins of all particles are in principle measurable, there are not necessarily any observables associated with the $z$-components of the \lq q-spins' of the anyons. Only $SU(2)_q$ invariant quantities, such as the total q-spins of groups of anyons, can be observables, or at any rate topologically protected observables. This can be traced back to the superselection rule that says that the total q-spin of all anyons together must be trivial. If it were possible to measure the $z$-components of every anyons' q-spin, then the state obtained would no longer be invariant under $SU(2)_q$. In fact, a similar rule would hold for confined particles in gauge theory and so for a better analogy, one may think of the $SU(2)_q$ invariance as begin closer to an $SU(2)$ gauge symmetry rather than spin. 

It was shown by Freedman {\it et al.}~\cite{Freedman:02a} that the
anyonic theories with $k\geq 3$, $k\neq 4$ are universal for quantum
computation. Hence, for those theories, the Bell violating states can be
obtained by topological braiding operations alone acting, for example, on
the fiducial state $\ket{\phi_0}$.
We now check to see if it is possible to generate a state which violates the
inequality for the $k=2$ case. The 
$SU(2)_2$ anyons are believed to exist in the
$\nu=5/2$ plateau of the fractional quantum Hall effect~\cite{Moore and Read}
up to charge factors that affect the Abelian part of braiding. 
They
come in three varieties, the vacuum, $1$, the fermion, $\psi$, and the
non-Abelian anyon, $\sigma$, that satisfy the non-trivial fusion rules,
$\sigma\times \sigma=1+\psi$, $\sigma\times\psi=\sigma$ and
$\psi\times\psi=1$. The counterclockwise exchange of two $\sigma$ particles,
which fuse to either $1$ or $\psi$, results in the matrix evolution
$R=1\oplus i$ expressed in the basis labeled by the fusion channels $\{1,\psi\}$. 
The state evolution produced by the exchange of particles with no immediate fusion channel is
found by employing the recoupling matrix $F$. Expressed in the basis
$\{\ket{x(\sigma,\sigma)}_A\ket{y(\sigma,\sigma)}_B; x,y\in
\{1,\psi \}\}$, we have the following representation of the generators of
the braid group $\mathcal{B}_6$
\begin{equation}
\begin{array}{lll}
B_{1}&=&e^{-i\frac{\pi}{4} \sigma^x}\otimes {\bf 1}_2,
B_{2}=e^{-i\frac{\pi}{4} \sigma^z}\otimes {\bf 1}_2,
B_{3}=e^{-i\frac{\pi}{4}\sigma^x\otimes \sigma^z},\\
B_{4}&=& {\bf 1}_2\otimes e^{-i\frac{\pi}{4} \sigma^x},
B_{5}={\bf 1}_2\otimes e^{-i\frac{\pi}{4} \sigma^z}
\end{array}
\end{equation}
where $B_j$ results from the exchange of $j$ and $j+1$ particles in a
counterclockwise manner. As a simple initial state we can consider
$\ket{\phi_0}$ that is produced from $(1,2)$, $(3,4)$ and $(5,6)$ pairs
created from the vacuum.


The braid group generators, $B_j$, are in the Clifford group, so we cannot
generate a dense set in $SU(4)$ by braiding alone\footnote{Complementing topological operations by some noisy non-topological operations, one can achieve universality~\cite{Bravyi:06}}. But can we still obtain Bell violating states? In Refs.~\cite{Spekkens:07,Wootters} a LHV model was introduced for a pair of qubits that exactly reproduces the set of allowed operations in the present model. There are two distinct configurations of shared vacuum pairs (up to relabeling of particles by Alice or Bob) both of which can be obtained from $\ket{\phi_0}$ by braiding.  Hence it is not possible to build Bell violating states starting out from three shared vacuum pairs using topologically protected operations alone.  
 
Despite the impossibility of producing a Bell violating state from $\ket{\phi_0}$ by topological gates, one can in fact straightforwardly obtain a maximally Bell violating state 
using non-topological gates\footnote{For an alternative construction of this kind, but using the more standard tensor product structure of $4$-anyon qubits, giving a total of 8 anyons, see~\cite{Zhang:07}.}.  Let us employ the non-Clifford gate $D=e^{-i\frac{\pi}{8}\sigma^z}\otimes {\bf 1}$.
This can be implemented by bringing the $2$ and $3$ $\sigma$ anyons nearby, thus
shifting the energy of the fermionic fusion
channel ~\cite{Lahtinen} such that a relative phase $e^{i\pi/4}$ is accumulated on that channel. From these operations one can build the controlled
phase gate in the following way $CP = e^{i\pi/4}
B_2B_1B_2B_3^{-1}B_2^{-1}B_1^{-1}B_5$. A simple searching algorithm provides us with the sequence that
produces $\ket{r(a_-)}=-CPB_3B_4DB_2B_3\ket{\phi_0}$, with $\langle
W\rangle=-2\sqrt{2}$, thus saturating the Tsirleson bound.

In fact, it is indeed possible to realize a maximally violating state without braiding at all.
Consider a state $\ket{\phi_0'}$ given by a distribution of singlet pairs on $(1,6)$, $(2,5)$ and $(3,4)$.  This state $\ket{\phi_0'}$
is related to the fiducial distribution of pairs by the following braid word
 $\ket{\phi'_0}=\frac{1}{\sqrt{2}}(\ket{0'}_A\ket{0}_B+\ket{1'}_A\ket{1}_B)=B_2^{-1}B_3^{-1}B_5B_4B_3B_2\ket{\phi_0}$.  This state would be maximally violating if we could measure in arbitrary local bases.  For our fixed measurement quorum $\ket{\phi'_0}$ is related to the maximally violating state by $e^{-i\frac{\pi}{8}\sigma^y}\otimes {\bf 1}\ket{\phi_0'}=\ket{r(a_+)}$ and hence it suffices to implement the local unitary $e^{-i\frac{\pi}{8}\sigma^y}=e^{-i\frac{\pi}{4}\sigma^z}e^{-i\frac{\pi}{8}\sigma^x}e^{i\frac{\pi}{4}\sigma^z}$ on Alice's side to obtain a Bell violation.  The $z$ rotation is simply achieved by bringing anyons $2$ and $3$ near each other as above, and similarly the $x$ rotation is performed by pushing $1$ and $2$ together.
Note that the maximal Bell violation in these two constructions actually saturates the Tsirelson inequality, making the $SU(2)_2$ case at the same time \lq maximally quantum mechanical' and \lq topologically classical'.

Let us turn now to Fibonacci anyons from the
$SO(3)_3$ theory. 
This is the theory obtained from $SU(2)_3$ but using only
integer spin particles: the vacuum $1$ and the
non-Abelian anyon $\tau$, with non-trivial fusion rule $\tau \times
\tau=1+\tau$. All particles are their own anti-particles and the quantum
dimensions are $d_1=1$ and $d_{\tau}=\phi\equiv(1+\sqrt{5})/2$. The
relevant recoupling matrix is
\[
F=F_{\tau\tau\tau}^{\tau}=\left(\begin{array}{cc}\phi^{-1} & \phi^{-1/2} \\
\phi^{-1/2} & -\phi^{-1}\end{array}\right)
\label{Ffib}
\]
expressed in the basis of $1$ and $\tau$. The dimension of the topological
Hilbert space of $m+1$ type $\tau$ anyons with total charge zero is $f_{m}$,
the $m$th Fibonacci number, hence there are five states in the fusion space.
These states can be decomposed into superpositions of products of local
basis states as in Eq. \ref{vects} where $\ket{0}=\ket{1(\tau,\tau)}$ and 
$\ket{1}=\ket{\tau(\tau,\tau)}$ and
$\ket{\phi_4}=\ket{\tau(\tau,1)}_A\ket{\tau(\tau,1)}_B$. The state $\ket{\phi_0}$ is the
state obtained by creating type $\tau$ particle anti-particle pairs on
$(1,2),(3,4),(5,6)$ out of the vacuum.   For the one parameter
family of states $\ket{r(a)}$ we find the same maximal violation as in Eq.
\ref{as} for $SU(2)_3$: $\langle W\rangle =\pm 2\sqrt{-7+4\sqrt{5}}\approx
\pm 2.7887$. This is not very surprising, considering that the $SU(2)_3$ theory is equivalent to the product of the Fibonacci theory and an Abelian theory with $\mathbb{Z}_2$ fusion rules (see for instance~\cite{Bonderson_thesis}).
The action under braiding is represented by the matrix
$R_{\tau\tau}=e^{i4\pi/5}\oplus e^{i7\pi/5}$
expressed in the basis $\{1,\tau\}$.   We obtain the following representation of the generators of the braid group $\mathcal{B}_6$ expressed in the basis
$\{\ket{0}\ket{0},\ket{0}\ket{1},\ket{1}\ket{0},\ket{1}\ket{1}\}\sqcup \ket{\phi_4}$:
\begin{equation}
\begin{array}{lll}
B_{1}&=&[FR_{\tau\tau}F^{-1}\otimes {\bf 1}_2]\oplus (e^{i7\pi/5})\\
B_{2}&=&[R_{\tau\tau}\otimes {\bf 1}_2]\oplus (e^{i7\pi/5})\\
B_{3}&=&O^{\dagger}[e^{i4\pi/5}\oplus e^{i7\pi/5}\oplus e^{i7\pi/5}\oplus\left(\begin{array}{cc}M_{1}^{1} & M_0^{1} \\ M_{1}^0 & M_0^0 \end{array}\right)]O\\
B_{4}&=&[{\bf 1}_2\otimes FR_{\tau\tau}F^{-1}]\oplus (e^{i7\pi/5})\\
B_{5}&=&[{\bf 1}_2\otimes R_{\tau\tau}]\oplus (e^{i7\pi/5})\\
\end{array}
\label{BraidFib}
\end{equation}
where $O$ maps the product basis to the basis $\{\ket{\phi_j}\}_{j=0}^4$, and $M=(F_{\tau\tau\tau}^{\tau})^{-1}R_{\tau\tau}F_{\tau\tau\tau}^{\tau}$.
A length 25 braid word produces a Bell violation:
$\ket{\Psi}=[B_3B_4^{-1}B_1^{-1}B_3^{-1}B_2^{-1}]^5\ket{\phi_0}$
with $\bra{\Psi}W\ket{\Psi} =2.5310$.

In the models above, measurements by
Alice and Bob had two outcomes for the two fusion products of the anyons.  
To accommodate more outcomes we can use higher dimensional Bell
witnesses~\cite{Collins:02}.  We demonstrate how this works in another anyonic model 
with excitations in one to one
correspondence with irreducible representations of a Hopf algebra, $D(G)$,
the quantum double of a finite group $G$~\cite{Bais:92,Propitius98}.
The particles
can carry electric and magnetic charge and are labeled by $\Pi^{[\alpha]}_{R(N_{[\alpha]})}$ where
$[\alpha]$ denotes a conjugacy class of $G$ which labels the magnetic charge, and $R(N_{[\alpha]})$ denotes a unitary irreducible representation $R$ of the centralizer of an element in the conjugacy class $[\alpha]$, which labels the electric charge.
The dimension of the carrier space for each irreducible representation, which equals the quantum dimension of the particle
$\Pi^{[\alpha]}_{R(N_{[\alpha]})}$ is $d^{[\alpha]}_{R(N_{[\alpha]})}=|[\alpha]||R(N_{[\alpha]})|$.
The quantum dimensions satisfy the sum rule $\sum (d^{[\alpha]}_{R(N_{[\alpha]}})^2=|G|^2$.
We focus on the simplest non-Abelian finite group, $S_3$, the group of
permutations on three objects. Elements of $S_3$ are organized into three
conjugacy classes:  $[e]=\{e\}$ the identity element, $[t]=\{t_0,t_1,t_2\}$
the transpositions, and $[c]=\{c_+,c_-\}$ the cyclic permutations. The $8$
irreducible representations for $D(S_3)$ are 
\begin{equation}
\begin{array}{lll}
&&\Pi^{[e]}_{R_1^+}\quad d=1 \quad {\rm (vacuum)}\\
&&\Pi^{[c]}_{\beta_0}, \Pi^{[t]}_{\gamma_0}\quad d=2,3 \quad {\rm (pure\ magnetic\ charges)} \\
&&\Pi^{[e]}_{R_1^-}, \Pi^{[e]}_{R_2}\quad d=1,2 \quad {\rm (pure\ electric\ charges)}\\
&&\Pi^{[c]}_{\beta_1}, \Pi^{[c]}_{\beta_2},\Pi^{[t]}_{\gamma_1}\quad d=2,2,3 \quad {\rm (dyonic\ combinations)}\\
\end{array}
\end{equation}

A complete derivation of the fusion rules for this model is given in \cite{Propitius:95}. In the toric code realization of these anyon models, the quantum dimensions $d^{[\alpha]}_{R(N_{[\alpha]})}=|[\alpha]||R(N_{[\alpha]})|$ actually count local degrees of freedom associated with the anyon. In the discrete gauge theory context, these degrees of freedom are also present in the description of the system, but some of them are gauge.  A single particle's electric charge and magnetic charge can always be measured locally (or at least within a region of size characteristic of the particles), by braiding with other locally prepared charge pairs and measuring the outcome of fusion of the pairs.  Truly non-local properties are contained in the fusion space.  To explore this we pick a fusion subalgebra of $D(S_3)$:  $\{\Pi^{[e]}_{R_1^+},\Pi^{[e]}_{R_1^-},\Pi^{[c]}_{\beta_0}\}$ which we label for convenience $\{1,\Lambda,\Phi\}$.  The non trivial fusion rules are
\[
\Lambda\times \Lambda=1,\quad \Lambda\times \Phi=\Phi,\quad \Phi\times \Phi=1+\Lambda+\Phi.
\]
These fusion rules are the same as the fusion rules for the representations of $S_3$ itself and also the same as the fusion rules of the integer spin sectors of $SU(2)_4$.
The particles are their own anti-particles.  The magnetic charge $\Phi$
with quantum dimension $2$ carries non-Abelian statistics and the fusion of
$n$ such particles gives:  $\Phi^{\times
n}=\frac{1}{3}(2^{n-1}+(-1)^n)(1+\Lambda)+\frac{1}{3}(2^{n}+(-1)^{n-1})\Phi$.
As before, we will work in the superselection sector with total trivial
charge.  The smallest number of particles in this sector that could hope to
violate a Bell inequality should have fusion space dimension $\geq 4$.  If
we are to pick measurement operators for Alice and Bob that measure total
charge on pairs of $\Phi$ particles and we want two non commuting operators
on each side then we require at least six particles in total. Exactly six
particles suffices, giving Hilbert space dimension eleven for the vacuum sector.

Either by using the representation theory of $D(S_3)$, or by solving the pentagon and hexagon equations directly, we find the following recoupling and braid matrices, expressed in the basis $\{1,\Lambda,\Phi\}$,
\[
F\equiv F_{\Phi\Phi\Phi}^{\Phi}=\left(\begin{array}{ccc}\frac{1}{2}& \frac{1}{2} &-\frac{1}{\sqrt{2}}  \\ \frac{1}{2} & \frac{1}{2} & \frac{1}{\sqrt{2}} \\ -\frac{1}{\sqrt{2}} &\frac{1}{\sqrt{2}}  & 0\end{array}\right);\ 
R\equiv R_{\Phi\Phi}= \left(\begin{array}{ccc}1&0&0 \\ 0 &-1&0 \\ 0&0 & 1\end{array}\right).
\]
We notice immediately that $R$ has eigenvalues $\pm 1$, so that we will end up with a representation of the permutation group when \lq braiding' the anyons. Nevertheless, these anyons are not bosons or fermions, since this representation is non-Abelian. The fact that we have a permutation group representation does signal the fact that braiding in this theory is not universal for quantum computation. This is in fact a general property of braiding in discrete gauge theories.

A basis of the eleven dimensional vacuum sector of the six anyon Hilbert space can be given in terms of superpositions of products of local basis states, as defined in figure~\ref{fig:3},
\[
\begin{array}{lll}
\{\sum_{y}F^y_1\ket{y(\Phi,\Phi)}_A\ket{1(\Phi,\Phi)}_B,
\sum_{y}F^y_{\Lambda}\ket{y(\Phi,\Phi)}_A\ket{1(\Phi,\Phi)}_B\\
\sum_{y}F^y_{\Phi}\ket{y(\Phi,\Phi)}_A\ket{1(\Phi,\Phi)}_B,
\sum_{y}F^y_1\ket{y(\Phi,\Phi)}_A\ket{\Lambda(\Phi,\Phi)}_B\\
\sum_{y}F^y_{\Lambda}\ket{y(\Phi,\Phi)}_A\ket{\Lambda(\Phi,\Phi)}_B,
\sum_{y}F^y_{\Phi}\ket{y(\Phi,\Phi)}_A\ket{\Lambda(\Phi,\Phi)}_B\\
\sum_{y}F^y_1\ket{y(\Phi,\Phi)}_A\ket{\Phi(\Phi,\Phi)}_B,
\sum_{y}F^y_{\Lambda}\ket{y(\Phi,\Phi)}_A\ket{\Phi(\Phi,\Phi)}_B\\
\sum_{y}F^y_{\Phi}\ket{y(\Phi,\Phi)}_A\ket{\Phi(\Phi,\Phi)}_B,
\ket{\Phi(\Phi,\Lambda)}_A\ket{\Phi(\Phi,\Lambda)}_B\\
\ket{\Phi(\Phi,1)}_A\ket{\Phi(\Phi,1)}_B\}=\{\ket{\phi_j}\}_{j=0}^{10}.
\end{array}
\]
The state $\ket{\phi_0}$ is obtained by creating type $\Phi$ particle anti-particle pairs on $(1,2),(3,4),(5,6)$ out of the vacuum. 
 Each such vacuum magnetic charge pair is written:
$\ket{\Phi,\Phi;(i,j)}=\frac{1}{\sqrt{2}}(\ket{c_+,c_-;(i,j)}+\ket{c_-,c_+;(i,j)})$.

Now in analogy to the cases studied for $SU(2)_k$, we could look for a Bell like inequality but using measurement operators with three outcomes.  Let Alice have one operator $\Upsilon^A_{1,2}$ which measures the outcome of total charge for particles $1$ and $2$ with outcomes $\{1,\Lambda,\Phi\}$ and another, non commuting operator, $\Upsilon^A_{2,3}$ that measures total charge for particles $2$ and $3$ with outcomes $\{1,\Lambda,\Phi\}$.  In other words, $\Upsilon^A_{1,2}$ is a measurement in the basis $\{(F_{\alpha\alpha\alpha}^{\beta})^{\dagger}\ket{y(\alpha,\beta)}_A\}$ with outcome $m^A_{1,2}=y\in\{1,\Lambda,\Phi\}$ and $\Upsilon^A_{2,3}$ is a measurement in the basis $\{\ket{y(\alpha,\beta)}_A\}$ with outcome $m^A_{2,3}=y$.  Similarly, let Bob have two measurement operators, $\Upsilon^B_{4,5}$ that measures in the basis $\{(F_{\alpha\alpha\alpha}^{\beta})^{\dagger}\ket{y(\alpha,\beta)}_B\}$ with outcome $m^B_{4,5}=y$, and $\Upsilon^B_{5,6}$ which measures onto the basis $\{\ket{y(\alpha,\beta)}_B\}$ with outcome $m^B_{5,6}=y$.  Now $(F_{\Phi\Phi\Phi}^{1})^y_x=\delta_{x,\Phi}\delta_{y,\Phi}=(F_{\Phi\Phi\Phi}^{\Lambda})^y_x$, so in the subspace of $\{\ket{\phi_9},\ket{\phi_{10}}\}$, the measurement operators all commute.  These states cannot yield a Bell violation and we can focus on the states in the $9$ dimensional orthogonal subspace which is isomorphic to the Hilbert space of two three dimensional particles (qutrits).

In Ref.~\cite{Collins:02} it was shown how to construct Bell inequalities for bipartite systems of equal but arbitrary finite dimension.  In particular for two qutrits the authors introduce the witness $I_3$ which for all LHV theories satisfies $|\langle I_3 \rangle| \leq 2$, whereas for quantum mechanical systems $|\langle I_3 \rangle| \leq 4$. To simplify notation let us introduce the projectors $\pi_{y}\equiv \ket{y(\Phi,\Phi)}\bra{y(\Phi,\Phi)}$ and $\tilde{\pi}_{y}\equiv F^{\dagger}\ket{y(\Phi,\Phi)}\bra{y(\Phi,\Phi)}F$.  For the quorum of observables above, the Bell witness $I_3$ is:
\begin{equation}
\begin{array}{lll}
I_3&=& \tilde{\pi}_{1}\otimes  \tilde{\pi}_{1} + \tilde{\pi}_{\Lambda}\otimes  \tilde{\pi}_{\Lambda} + \tilde{\pi}_{\Phi}\otimes  \tilde{\pi}_{\Phi}+ \pi_{\Phi}\otimes  \tilde{\pi}_{1} + \pi_{1}\otimes  \tilde{\pi}_{\Lambda} \\
&&+ \pi_{\Lambda}\otimes  \tilde{\pi}_{\Phi}
+  \pi_{1}\otimes  \pi_{1} + \pi_{\Lambda}\otimes  \pi_{\Lambda} + \pi_{\Phi}\otimes  \pi_{\Phi}
+  \tilde{\pi}_{1}\otimes  \pi_{1} \\
&&+ \tilde{\pi}_{\Lambda}\otimes  \pi_{\Lambda} + \tilde{\pi}_{\Phi}\otimes \pi_{\Phi}
-  \tilde{\pi}_{1}\otimes  \tilde{\pi}_{\Lambda} + \tilde{\pi}_{\Lambda}\otimes  \tilde{\pi}_{\Phi} + \tilde{\pi}_{\Phi}\otimes  \tilde{\pi}_{1}\\
&&-\pi_{1}\otimes  \tilde{\pi}_{1}+ \pi_{\Lambda}\otimes  \tilde{\pi}_{\Lambda} + \pi_{\Phi}\otimes  \tilde{\pi}_{\Phi}- \pi_{1}\otimes  \pi_{\Lambda} + \pi_{\Lambda}\otimes  \pi_{\Phi} \\
&&+ \pi_{\Phi}\otimes  \pi_{1}
 - \tilde{\pi}_{\Lambda}\otimes  \pi_{1} + \tilde{\pi}_{\Phi}\otimes  \pi_{\Lambda} + \tilde{\pi}_{1}\otimes  \pi_{\Phi}\\
 &&
+2 \ket{\Phi(\Phi,\Lambda)}\bra{\Phi(\Phi,\Lambda)}\otimes \ket{\Phi(\Phi,\Lambda)}\bra{\Phi(\Phi,\Lambda)}\\
&&+2 \ket{\Phi(\Phi,1)}\bra{\Phi(\Phi,1)}\otimes \ket{\Phi(\Phi,1)}\bra{\Phi(\Phi,1)}
\end{array}
\end{equation}
The state with the largest violation has $\langle I_3\rangle=-2.5216$.

We obtain the following representation of the generators for $\mathcal{B}_6$ expressed in the basis
$\{\ket{x(\Phi,\Phi)}_A\ket{y(\Phi,\Phi)}_B; x,y\in \{1,\Lambda,\Phi \}\}\sqcup \ket{\Phi(\Phi,\Lambda)}_A\ket{\Phi(\Phi,\Lambda)}_B\sqcup \ket{\Phi(\Phi,1)}_A\ket{\Phi(\Phi,1)}_B\} $:
\begin{equation}
\begin{array}{lll}
B_{1}&=&[FRF^{-1}\otimes {\bf 1}_3]\oplus R^{\Phi}_{\Phi\Phi} \oplus R^{\Phi}_{\Phi\Phi},B_{2}=[R \otimes {\bf 1}_3]\oplus R^{\Phi}_{\Phi\Phi} \oplus R^{\Phi}_{\Phi\Phi} \\
B_{3}&=&O^{\dagger}\Bigg[R^{1}_{\Phi\Phi}\oplus R^{\Lambda}_{\Phi\Phi}\oplus R^{\Phi}_{\Phi\Phi}\oplus R^{\Lambda}_{\Phi\Phi} \oplus R^{1}_{\Phi\Phi}\oplus R^{\Phi}_{\Phi\Phi}\oplus\\
&&R^{\Phi}_{\Phi\Phi}\oplus R^{\Phi}_{\Phi\Phi}\oplus \left(\begin{array}{ccc}M^{\Phi}_{\Phi} & M^{\Phi}_{\Lambda} & M^{\Phi}_{1} \\M^{\Lambda}_{\Phi} & M^{\Lambda}_{\Lambda} & M^{\Lambda}_{1} \\M^{1}_{\Phi} & M^{1}_{\Lambda} & M^{1}_{1}\end{array}\right)\Bigg]    O\\
B_{4}&=&[{\bf 1}_3\otimes FRF^{-1}]\oplus R^{\Phi}_{\Phi\Phi}\oplus R^{\Phi}_{\Phi\Phi},
B_{5}=[{\bf 1}_3\otimes R]\oplus R^{\Phi}_{\Phi\Phi} \oplus R^{\Phi}_{\Phi\Phi}\\
\end{array}
\label{BraidQD}
\end{equation}
where
$O$ maps the product basis to the basis $\{\ket{\phi_j}\}_{j=0}^{10}$, and $M=F^{-1}RF$.
Here $B_j^2={\bf 1}_{11} \forall j$, so we have the permutation group $S_6$, as mentioned before.  We compute the action on states consisting of vacuum magnetic charge pairs.  The state $\ket{\phi_0}=\ket{\Phi,\Phi;(1,2)}\ket{\Phi,\Phi;(3,4)}\ket{\Phi,\Phi;(5,6)}$ is the fiducial state, and the other distinct configuration of vacuum magnetic charge pairs is $\ket{\Phi,\Phi;(1,4)}\ket{\Phi,\Phi;(2,5)}\ket{\Phi,\Phi;(3,6)}=B_3B_4B_2B_1\ket{\phi_0}$, hence it suffices to consider the orbit of $\ket{\phi_0}$.  An exhaustive search through $6!=720$ braid words corresponding to all distinct permutations in $S_6$ finds that, while $\langle I_3\rangle$ is not constant under braiding, we do find that in all cases $|\langle I_3\rangle |\leq 2$. Hence we require some operation beyond braiding to produce a violation of LHV under our protocol. 
Even if we restrict to non-topologically protected operations that just involve interacting pairs of particles, we can indeed produce a Bell violating state.  Consider the family of states $\ket{\phi'}=D_{3,4}(\alpha_1,\alpha_2)D_{1,2}(\alpha_3,\alpha_4)D_{2,3}(\alpha_5,\alpha_6)B_1B_5B_3B_2B_3B_4\ket{\phi_0}$ where 
$D_{i,j}(\alpha,\beta)$ is the non-topologically protected gate obtained by bringing anyons $i$ and $j$ of type $\Phi$ nearby each other and allowing them to interact for a time such that the fusion channel $\Phi\times\Phi\rightarrow \Lambda$ accumulates a phase $e^{i\alpha}$ and the fusion channel $\Phi\times\Phi\rightarrow \Phi$ accumulates a phase $e^{i\beta}$.  Optimizing $|\langle I_3\rangle|$ over the interaction phases, we find a violation $\langle \phi' |I_3|\phi'\rangle=2.0512$ for the angles:
$\alpha_1=0.7943,\alpha_2=0.3989,\alpha_3=3.5531,\alpha_4=0.9257,\alpha_5=-0.8525,\alpha_6=0.1036$.  No systematic attempt was made to optimize the violation over other braid words and it likely stronger violations could be found.

We have described a protocol to reveal non-locality in several classes of non-Abelian anyonic theories.  The need for at least six anyons shared between two parties arises because each party needs three anyons in order to have two non-commuting topologically protected observables.
It is possible this could be reduced using a shared resource which fixes a common gauge, akin to using a shared reference frames to reveal non-locality in mode entanglement with bosons~\cite{BRS}. The size of the maximum violation depends on the recoupling matrices $F$ and the ability to generate Bell violating states beginning from 3 vacuum charge pairs depends on the power of the braiding operations.  It is intriguing to ask whether one could find intermediate anyonic theories which have the power to generate Bell violating states by topologically protected gates, but are not universal for topological quantum computation.  Finally, it would be very interesting to have an experimental demonstration of (some of) the schemes presented here. We believe that the required effort would not be significantly higher than that necessary to perform non-abelian interferometry.

{\it Acknowledgements}  S.I. acknowledges support from the Generalitat de Catalunya, MEC (Spain), and the European project QAP.

 \end{document}